\documentclass[manuscript]{acmart}

\AtBeginDocument{%
  }

\setcopyright{none}
\renewcommand\footnotetextcopyrightpermission[1]{}
\usepackage{longtable}
\usepackage{multirow}

\begin{document}

\title{Understanding Creative Design Practices among Data Artists}

\author{Tianwei Ma}
\email{tianwei-ma@uiowa.edu}
\orcid{0009-0009-9819-3982}
\affiliation{%
  \institution{University of Iowa}
  \city{Iowa City}
  \country{USA}
}

\author{Anna Offenwanger}
\email{offenwanger@uvic.ca}
\orcid{0000-0001-9560-201X}
\affiliation{%
  \institution{University of Victoria}
  \city{Victoria}
  \country{Canada}
}

\author{Naimul Hoque}
\email{naimul-hoque@uiowa.edu}
\orcid{0000-0003-0878-501X}
\affiliation{%
  \institution{University of Iowa}
  \city{Iowa City}
  \country{USA}
}

\renewcommand{\shortauthors}{Ma et al.}

\begin{abstract}
Creative and artistic data visualizations communicate stories and invite engagement, yet how designers develop their expressive forms remains poorly understood. We investigate the design process of data artists through three complementary studies: an analysis of 40 public project accounts, artifact-anchored interviews with seven experienced data artists, and a design task-based study with eight data artists. We find that stories and visual forms develop together through data exploration, reference adaptation, sketching, and prototyping. Inspiration comes from data, existing work, everyday imagery, and personal experience. Sketches help develop mappings, while prototypes with real data can reshape representations and intended stories. Designers consider multiple possibilities but typically develop one direction at a time, partly because producing alternatives is costly. Their choices balance meaning, visual appeal, readability, and feasibility. These findings inform tools connecting stories, references, and real data, including AI assistance that supports testing and revising alternatives while preserving designers’ creative judgment.
\end{abstract}



\maketitle
\section{Introduction}
Creative and artistic data visualization has become an established area of practice. IEEE VIS has hosted a dedicated Arts Program since 2013, which accepted more than two hundred artworks in its first decade~\cite{lan2025beautiful}. The Information is Beautiful Awards, founded in 2012, judge work in categories that explicitly admit information art, sculpture, and installation, and drew nine hundred submissions from more than fifty countries in 2023.\footnote{\url{https://www.datavisualizationsociety.org/news/2023-information-is-beautiful-awards-winners}} The work circulates well beyond the visualization community. \emph{Dear Data}~\cite{lupi2016deardata}, a year-long exchange of hand-drawn postcards between Giorgia Lupi and Stefanie Posavec, was acquired by the Museum of Modern Art for its permanent collection in 2016. In 2022, Lupi received the National Design Award in communication design from the Cooper Hewitt, Smithsonian Design Museum. What Vi\'egas and Wattenberg could describe in 2007 through a handful of examples~\cite{viegas2007artistic} now supports careers, commissions, exhibitions, and a professional society. Despite this growing body of application and interest, we still know very little about how these creative and artistic visualizations are made.  

Visualization research has produced methodologies for conducting design work~\cite{DBLP:journals/tvcg/SedlmairMM12, DBLP:journals/tvcg/McKennaMAM14, DBLP:journals/tvcg/RobertsHR16}. Empirical studies have examined how designers frame problems~\cite{parsons2025beyond}, find and use examples~\cite{bako2022understanding, bako2024unveiling}, seek inspiration~\cite{viz_inspiration}, sketch~\cite{walny2015sketching}, work with real data~\cite{bigelow2014reflections}, and hand designs to developers~\cite{walny2020data}. Focusing specifically on artistic data visualization, Lan et al.~\cite{lan2025beautiful} analyzed 220 artworks and interviewed twelve data artists to characterize design features and understand creative practices. They identified input-driven workflows, in which artists begin by collecting and analyzing data, and output-driven workflows, in which artists first envision a desired outcome and then use data to realize it. These findings describe broad approaches to creation, but how initial directions develop through successive design attempts remains less understood. In particular, we need a more detailed account of how data artists draw on inspiration, test and revise candidate representations, and respond to difficulties as their designs develop.

Our study characterizes ideation in creative and artistic data visualization: the trajectory through which designers move from an initial dataset or concept toward a complete visual format. We examine this process through three complementary studies. Study~I analyzes 40 publicly documented projects, providing a broad view across designers, project contexts, and output formats. Study~II uses artifact-anchored interviews with seven experienced practitioners to elicit details about inspirations, decisions, alternatives, and difficulties that may not be talked detail in written accounts. Study~III observes eight designers developing new work across a live session and a period of independent work, allowing us to examine ideation as it unfolds rather than only in retrospect. Together, these studies let us examine the process across breadth, contextual detail, and temporal immediacy.

Across these perspectives, we find that creative and artistic visualization follows a flexible, iterative trajectory. Designers develop stories through repeated exploration of what the data can support. Visual inspiration comes from both the data itself and external sources, including existing work, everyday imagery, and personal experience. Designers adapt references for both meaning and appearance, using sketches to develop mappings and prototypes with real data to test and reshape ideas. Although multiple possibilities are considered, development usually centers on one direction at a time, partly because producing alternatives is costly. Early difficulties more often prompt reframing or abandonment, while later ones lead to adjustments within an established direction. Decisions to continue or stop reflect designers’ judgments of meaning, visual appeal, readability, and feasibility. In the design task, AI supported exploration, prototyping, and critique while designers retained control over creative decisions.  We conclude by deriving implications for tools that support this process, including opportunities for AI assistance that preserves designers’ creative judgment.

\section{Related Work}
This work relates to the creative practices and visualization design process. We discuss these areas below.

\subsection{Creative Visualization Design}
 
  It is difficult to provide a concrete definition of creative  visualization design since creativity itself is an abstract concept~\cite{ma2026characterizing}. In early 2000, Shneiderman argued that visualizing data is a creative task~\cite{DBLP:journals/cacm/Shneiderman07a}. It is most likely that Shneiderman was referring to standard charts (e.g., bar and pie charts) as a medium for creative exploration. However, colloquial references to creative visualization over time have moved towards non-traditional and personalized charts (e.g., infographics, pictographs, unusual layouts, etc.), which we also adopt in this paper. This is exemplified by the innovation of numerous authoring tools and techniques proposed for creating personalized and unorthodox charts~\cite{DBLP:conf/chi/RomatRHDAH20, schroeder2015visualization, DBLP:journals/tvcg/HuronVF13, DBLP:conf/chi/ZhangSBC20}. For example, Xia et al.~\cite{xia2018dataink} proposed DataInk, a sketching-based authoring tool that allows users to sketch glyphs with pen and touch interactions and bind the glyphs to actual data, enabling data-driven creative visualizations. Coelho and Mueller proposed the idea of Infomages~\cite{DBLP:journals/cgf/CoelhoM20}, a chart embedded into an image that represents the theme and context of the data. The authors also proposed an authoring tool that can help designers find thematic images from the web and embed them with data. While these tools and methods are shown to be effective as research prototypes and techniques, we have limited knowledge about how creative practitioners, such as data artists, design these creative visualizations. In a recent study, Ma et al.~\cite{ma2026characterizing} found that the design process and context, including the type of data, design goals, design inspirations, and organizational limitations, play a key role in the design of creative visualization. In this work, we aim to understand how these factors affect the design process of data artists. Our work has the potential to inform future visualization tools and techniques that are better aligned for creative workflows. A closely related work is the study by Lan et al.~\cite{lan2025beautiful}. The authors analyzed 200 data arts and conducted interviews with twelve data artists. The paper analyzed the final artifact while we investigate the design process that leads to the artifact. We provide step-by-step accounts of the design process followed by data artists based on three complementary studies and identify nuanced design practices that are not present in the study presented by Lan et al.

\subsection{Visualization Design Process}
Existing literature on visualization design process is relevant to this work because of our focus on design process.
Prior literature in visualization provides a number of systematic processes aimed at supporting design practice and teaching. These include proposing new design processes and validating them with user studies and deployments~\cite{DBLP:journals/tvcg/RobertsHR16, DBLP:journals/tvcg/SedlmairMM12, DBLP:journals/tvcg/KerznerGDJM19, DBLP:journals/tvcg/McKennaMAM14}. Perhaps the most well-known among these processes is the ``Design Study Methodology'' by Sedlmair et al~\cite{DBLP:journals/tvcg/SedlmairMM12}. The authors proposed 9 distinct phases to guide visualization design: \textit{Learn, Winnow, Cast, Discover, Design, Implement, Deploy, Reflect,} and \textit{Write}. Syeda et al.~\cite{DBLP:conf/chi/SyedaMRBB20} converted this methodology to a time-bound (i.e., lite) version, enabling instructors to adapt the methodology in a 14-week university course on data visualization. 
Some design frameworks focus on outlining low-level design tasks that designers and researchers can perform systematically to achieve a desirable solution~\cite{DBLP:journals/tvcg/GoodwinDJDDDKSW13, DBLP:conf/chi/MendezHN17, DBLP:journals/tvcg/McKennaMAM14, DBLP:journals/tvcg/KerznerGDJM19, DBLP:journals/tvcg/RobertsRJH18}. For example, McKenna et al.~\cite{DBLP:journals/tvcg/McKennaMAM14} proposed the Design Activity Framework, where each activity has three components: motivation, outcomes, and methods that are divergent, convergent, or both in nature. Authors also proposed four different activities that directly map to the nested visualization model~\cite{munzner2009nested}. The five design-sheet methodology~\cite{DBLP:journals/tvcg/RobertsHR16} proposes a predefined structure for five sheets where learners can creatively explore ideas through sketching.  

Another thread of research investigated actual design practices among users~\cite{design_fixtation, viz_inspiration, writing_rudders, daneshzand2025designing}. Parsons~\cite{design_practice} interviewed 20 data visualization practitioners to understand the methods and steps they take to design visualizations. The findings indicate that real-world designers often do not highly systematic, prescribed processes (like the ones discussed above), but instead rely on situated knowledge, precedence, experience, and judgment. Baigelenov et al.~\cite{viz_inspiration} investigated sources of inspiration for visualization design, outlining the role of existing visualizations, real-world phenomena, and personal experiences in inspiring new designs. Stokes et al.~\cite{writing_rudders} found that few designers use ``writing'' as an externalizing method in the early-stages of visualization design. However, in a followup study, authors found that writing, especially when done structurally to produce questions and future takeaways, can be useful to designers.

A key gap in the literature is the extent of alignment between academic visualization design frameworks and the actual design processes of users who are known for being eccentric and creative (e.g., data artists). This bidirectional mapping can be useful for both general visualization designers and creative professionals. Designers can learn about fostering creativity, while creative professionals can learn how to systematically approach visualization concepts like marks, channels, and layouts. Similarly, the knowledge can inform future research about merging the two approaches to visualization designs.

\section{Methodology}
We approached our research question through three complementary studies designed to provide different forms of access to creative and artistic visualization practice. Each study was designed to make visible aspects of the process that are difficult to capture through the other approaches, collectively spanning the breadth of creative practice, the situated details elicited through reflection, and the unfolding of ideation over time. The studies should therefore be understood as complementary lenses on the same phenomenon rather than as independent investigations or repeated validations.

To provide a broad view across creative visualization practices, study~I analyses a systematically collected set of 40 first-person accounts to identify recurring patterns in ideation, development, and challenges. While published accounts provide a wide body of evidence for ideation patterns, their reliance on how much authors choose to share may leave less salient decisions, exploratory attempts, or abandoned directions undocumented, resulting in incomplete accounts of the design process. In order to capture a more wholistic picture and specifically probe on abandoned directions and exploratory attempts, study II involved artifact-anchored interviews with seven practitioners with substantial professional experience in creative or artistic data visualization. Rather than asking participants to reconstruct an entire project chronologically, we used targeted questions and participants' own project materials to probe particular moments, decisions, alternatives, and difficulties. The conversational setting served as an elicitation mechanism, allowing participants to recover details that might not have surfaced through unprompted recollection. Yet these interviews remain retrospective and are still shaped by memory and later interpretation.

In order to capture information which is lost from retrospective accounts, we conducted Study~III---a design task---to address this temporal limitation by observing ideation as it unfolds. Eight participants with varied backgrounds in visualization, design, and art developed a new concept across a live ideation session and a period of independent work, documenting notable events with prompt cards and later walking us through what they had produced. By speaking with participants immediately we aim to capture emerging ideas, changes of direction, abandoned possibilities and a much more verbose collection of intermediate artifaces and sketches. It also allows us to examine whether patterns identified in documented and retrospective accounts appear in situated activity. However, the study is necessarily bounded: participants were able to develop their ideas through independent work, but were not required to carry the work through to a complete professional project. We intentionally set this scope to limit participant commitment.

Together, these studies provide complementary evidence at different levels of breadth, articulation, and temporal immediacy, enabling us to characterize the phenomenon from multiple perspectives.

\subsection{Study I: Public Accounts Analysis}
A rich primary source of data about creative visualization designers' processes are first hand accounts of visualization projects that appear in practitioner magazines, in conference pictorials, and on designers' own websites. Such accounts offer an unusually close-to-the-work record of the design process: authors often document a project soon after completing it, and many include background materials and intermediate artifacts such as sketches, mockups, and prototypes. These materials therefore preserve details and aspects of sequence that may be difficult to recover from later reflection.

\begin{table}[t]
\centering
\caption{Coding scheme.}
\Description{Two columns list 11 top-level codes and example sub-codes, with brief explanations of each category.}
\label{tab:S1_codebook}
\small
\begin{tabular}{@{}lp{0.75\columnwidth}@{}}
\toprule
\textbf{Top-level Code} & \textbf{Sub-level Code} \\
\midrule

& \textit{What the visualization format is. Project attribute, not an event.} \\
\cmidrule(l){2-2}
Output format &
Static 2D, 3D on screen, Physical (non-interactive), Physical (interactive), etc. \\
\midrule

& \textit{What the project is for. Project attribute, not an event.} \\
\cmidrule(l){2-2}
Overall visualization goal &
Evoke attention and discussion, Engaging to help gain insights, Personal interest, etc. \\
\midrule

& \textit{Obtaining the dataset. An event} \\
\cmidrule(l){2-2}
Collect Data & 
Public, Clients, Self-collected/constructed \\
\midrule

& \textit{Working with the data in hand to understand it or make it usable. An event} \\
\cmidrule(l){2-2}
Explore data & 
Make basic charts, Data selection, Transformed, Cleaning / preparation, etc. \\
\midrule

& \textit{Where an idea comes from, when the source is inside the project. An event} \\
\cmidrule(l){2-2}
Inspiration & 
Brainstorming, Eureka moment, Physical properties, Happy little mistakes, etc. \\
\midrule

& \textit{Where an idea comes from, when the source is outside the project. An event} \\
\cmidrule(l){2-2}
External inspiration & 
Metaphor, Others' work \\
\midrule

& \textit{Seeking feedback or information outside the creating unit. An event} \\
\cmidrule(l){2-2}
External input & 
Expert, Clients, Seek out knowledge or skills, Friends/families/colleagues/public \\
\midrule

& \textit{A proposal for what the work looks like, and the form it is externalized in. An event} \\
\cmidrule(l){2-2}
Visual idea & 
Mockups, Sketching, Described in words, Physical prototype \\
\midrule

& \textit{A point where the author reports that something does not work or is in the way. An event} \\
\cmidrule(l){2-2}
Blockers & 
Bad readability, Material or fabrication limitation, Skill or knowledge gap, No visual direction, etc. \\
\midrule

& \textit{What the author does in response to a blocker. An event} \\
\cmidrule(l){2-2}
Get unblock & 
Put aside, Reframe the problem, Brainstorming, Change the material or technique, Revert, etc. \\
\midrule

& \textit{Work on a design already established. An event} \\
\cmidrule(l){2-2}
Iterations/refinement & 
Mapping, Visual, Technical implementation, Material or medium choice, Explanatory aids, etc. \\

\bottomrule
\end{tabular}
\end{table}

\subsubsection{Corpus Sources}
\label{sec:corpus}

\paragraph{Sources}
In order to source accounts of visualization design processes, we sought out venues that publish accounts of visualization design processes where the articles can be enumerated to support systematic data collection and screening. We selected two suitable venues: VISAP Pictorials, a submission type within the IEEE VIS Arts Program \cite{visap2025}, and the Data Visualization Society Nightingale publication, where they regularly publishes behind-the-scenes pieces written by the designers of individual works, which has two achieves, the current website (July 2021 - present)\cite{nightingale2021site} and the previous medium publication (2019 - 2021)\cite{nightingale2019medium}. We also considered the Information is Beautiful awards\cite{iiba2026showcase} and other gallery-type collections, but concluded that the majority of these kinds of articles do not describe the visualization design process only the final visualization, so these venue were rejected. Instead, we supplemented the corpus with relevant articles found on individuals' websites and in the published book \textit{Data Sketches}~\cite{bremer2021datasketches} that we are aware of and come across in the course of research.

Our initial corpus consisted of all 37 pictorials published at VISAP between 2018 and 2025, 743 articles from the current Nightingale site, 596 from the Nightingale medium site, and 9 articles from designers' websites and books.

\paragraph{Inclusion Criteria}
We screened at two levels, applying one set of criteria to the work itself and another to the article describing it.
At the \textit{work} level, the piece had to be (a) a data visualization, and (b) be creative or artistic rather than analytic. We treated a work as creative if it satisfied at least one of three signals: it used a custom visual form rather than standard chart types; the author stated an artistic, emotional, or narrative intent; or it was produced for an art context such as an exhibition, a commission, or an art award. Dashboards, business reporting, and works assembled from conventional chart types were excluded.

At the \textit{article} level, we required four things. (a) The article had to be written in the first person by the designer or a member of the project team, so that it reports the process from inside rather than describing it from outside. (b) It had to be anchored to one identifiable work, or to a bounded series; general reflections on practice, career retrospectives, and methodology posts were excluded. (c) It had to cover the ideation trajectory, meaning that it showed the designer moving from a starting point---a dataset, a brief idea, or a question---toward a final visual concept; articles that only explained implementation, or only described the finished piece, were excluded. (d) The write-up had to have an identifiable temporal order, so that its steps could be placed on a timeline; articles that discussed a project thematically, without a recoverable sequence, were excluded. We restricted the corpus to English-language text that is publicly available.

As our interest is in the field rather than in individual practice, and a single designer's projects are likely to follow similar paths, we retained at most two accounts per author. If an author had more than two eligible articles, we kept the two that described the author's process in the most detail.

\begin{figure*}[t]
  \centering
  \includegraphics[width=\textwidth]{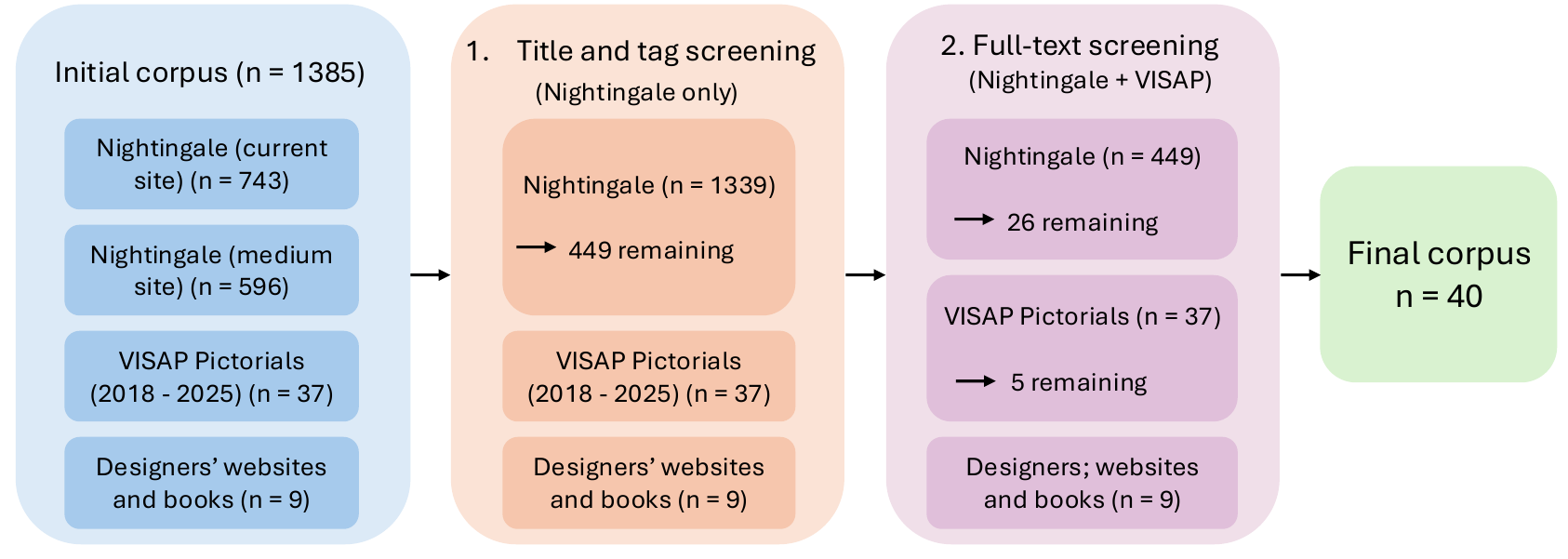}
  \caption{\textbf{Corpus construction and filtering steps for coding.}}
  \Description{A left-to-right screening flowchart.}
  \label{fig:corpus_filter}
\end{figure*}

\paragraph{Screening}
We first made a coarse pass over titles and tags to remove articles that could not be work-specific process accounts: eg, society and community news, tool tutorials, etc. We then read the remaining 449 articles against the criteria above. Screening was carried out by one author, who flagged uncertain cases; another author reviewed all flagged articles, and the two resolved them by discussion. The final corpus comprises 40 articles by 37 authors---5 from VISAP, 26 from Nightingale, and 9 from designers' websites and books---published between 2018 and 2025. Figure~\ref{fig:corpus_filter} reports counts at each stage of collection and screening. The full list of articles with links to the source can be found in the supplementary materials.

\subsubsection{Codebook Development}
Three authors independently performed bottom-up coding on a set of five accounts and compared results over several rounds of discussion. These rounds settled the structure of the codebook. Two authors then coded a further five accounts under this structure, and their comparison produced the initial codebook.

The remaining corpus was divided among the authors, with each article coded by one author. Coding and codebook development continued together: authors raised new or ambiguous cases as they arose, and sub-codes were added, split, or reworded by agreement. Whenever the codebook changed, previously coded accounts were revisited, so that all accounts were coded against the final version. Coding was carried out in a shared spreadsheet.

\subsubsection{Coding Scheme}
The codebook contains 11 top-level codes and 82 sub-codes. Table~\ref{tab:S1_codebook} contains all 11 top-level codes with example subcodes. The full codebook with descriptions is in Appendix~\ref{app:codebook}. Three conventions govern how codes are assigned. First, a passage may carry more than one sub-code when several apply at once; multiple sub-codes are assigned within one top-level code. Second, for narratives with out-of-order events, we code events by their actual order, not their textual order. Third, an event's top-level code records the role an action plays, so the same activities can appear under more than one top-level code: brainstorming, for instance, falls under \textit{Inspiration} where it arises on its own and under \textit{Get unblock} where it solves a stated blocker.

\subsubsection{Analysis of Process Sequences}
\label{sec:study1-analysis}

Our analysis uses the coded event sequence of each project to characterize how designers move through the ideation process and how they respond when difficulties arise. Coding produced 40 event chains, one per project, holding 682 sequential events. We characterize what activities make up these processes, when they tend to occur, and how designers move between them. We also consider how designers respond to difficulties and whether responses differ across different types or points of difficulty.

To compare chains, we use two measures to support these analyses. To describe \textit{how much} an activity occurs, we use both event counts and \textit{prevalence}. Event counts measure how many coded events belong to a code, while prevalence measures how many of the 40 projects contain that code at least once. We report both because the accounts vary in length: a longer account can contain more events simply because it provides more detail, whereas prevalence indicates how widespread an activity is across projects. To describe \textit{where} an activity occurs in a process, we use \textit{normalised position}. Each event is assigned a position from 0 to 1 based on its location within its own event sequence, so that positions can be compared across projects with different numbers of events. We report median positions across projects.

We examine the event sequences from four complementary perspectives. \textit{Composition} describes which activities make up the process and how widespread they are across projects. \textit{Position} describes where activities tend to occur within the process. \textit{Transitions} examine which activities tend to follow one another. \textit{Routines} identify recurring multi-step patterns across projects using PrefixSpan~\cite{pei2001prefixspan}. We then focus on the relationship between difficulties and the actions used to address them. Each \textit{Blocker} is linked to the \textit{Get unblock} that resolves it, even when other events occur in between. This helps us examine which activities are used to address different difficulties and whether the response changes depending on when or what type of difficulty arises.

Finally, we test whether these process patterns differ across project types. We compare whole trajectories across output formats and stated visualization goals using PERMANOVA~\cite{anderson2001permanova}, and compare individual codes using Fisher's exact test on prevalence with Benjamini--Hochberg correction~\cite{benjamini1995controlling}.

\subsection{Study II: Artifact-Anchored Interviews}

\begin{table*}[t]
\centering
\caption{\textbf{Study II participant demographics and practice.} We recruited 7 participants with diverse visualization expertise.}
\Description{Seven rows describe participants I1 to I7 by sex, age range, country, and artistic practice.}
\label{tab:S2_designers}
\small
\begin{tabular}{@{}llllp{0.63\textwidth}@{}}
\toprule
\textbf{ID} & \textbf{Sex} & \textbf{Age} & \textbf{Country} & \textbf{Practice} \\
\midrule
I1 & Male & 35--44 & US & Printed posters, some animated, mostly from public data \\
I2 & Female & 25--34 & US & 2D pieces mostly from personal data \\
I3 & Female & 25--34 & UK & 2D pieces from personal data \\
I4 & Female & 35--44 & Russia & 2D pieces and physical objects, from public and personal data \\
I5 & Male & 35--44 & France & Interactive installations and large-scale prints, from commissioned data \\
I6 & Female & 35--44 & US & 2D pieces from public data \\
I7 & Female & 45--54 & Switzerland & 2D pieces from public data \\
\bottomrule
\end{tabular}
\end{table*}

\subsubsection{Participants}
We interviewed seven experienced data artists (Table~\ref{tab:S2_designers}). To take part, a designer had to self-identify as a data visualization designer and maintain a public portfolio containing several works that meet the definition of creative or artistic data visualization and have at least five years of professional practice in it. We recruited by targeted email. Kirell Benzi (I5)\footnote{\url{https://kirellbenzi.com/}} is identified with his consent; the other interviewees are referred to by participant IDs.

\subsubsection{Procedure}
We used stimulated recall~\cite{bloom1953thought}: rather than asking designers to describe their practice in general, we anchored each interview to one to two completed projects of their own choosing and to the material they had produced while making it. Ahead of the interview we asked participants to gather whatever they still had from the projects---the dataset, sketches, mockups, prototypes, or notes---and to have it available during the conversation.

Each interview was a single session held over Zoom, lasting around 60 minutes. Interviews followed a template covering questions of where the inspiration came from, the main difficulties in the project and the points at which progress stalled, how ideas were externalised along the way, why the piece took the form it did and what other forms were considered, how they would approach a project they had never worked on, and an ideal tool that would help them during the ideation process. The interviews were recorded with participant consent on video and audio. Participant residing in the U.S. received \$35 for their time. We were not able to compensate participants outside the U.S. (4 out of 7) for organizational constraints. Their participation was voluntary.

\subsubsection{Analysis}
We analysed the interviews using reflexive thematic analysis~\cite{braun2019reflecting}. We transcribed the recordings with Zoom's automatic transcription and corrected each transcript shortly after the interview when the conversation was still fresh. Two authors read all seven transcripts and coded them separately, working from the participants' own wording rather than from a prepared scheme. They then went through their codes together, discussed where their readings differed, and grouped related codes into candidate themes over several rounds. We developed the themes from the interviews alone, without applying the codebook from Study~I. Once the themes were stable, we compared each of them with the corresponding result from Study~I and noted whether the two agreed, or whether the interviews raised something the corpus does not contain.

\begin{figure*}[t]
  \centering
  \includegraphics[width=\textwidth]{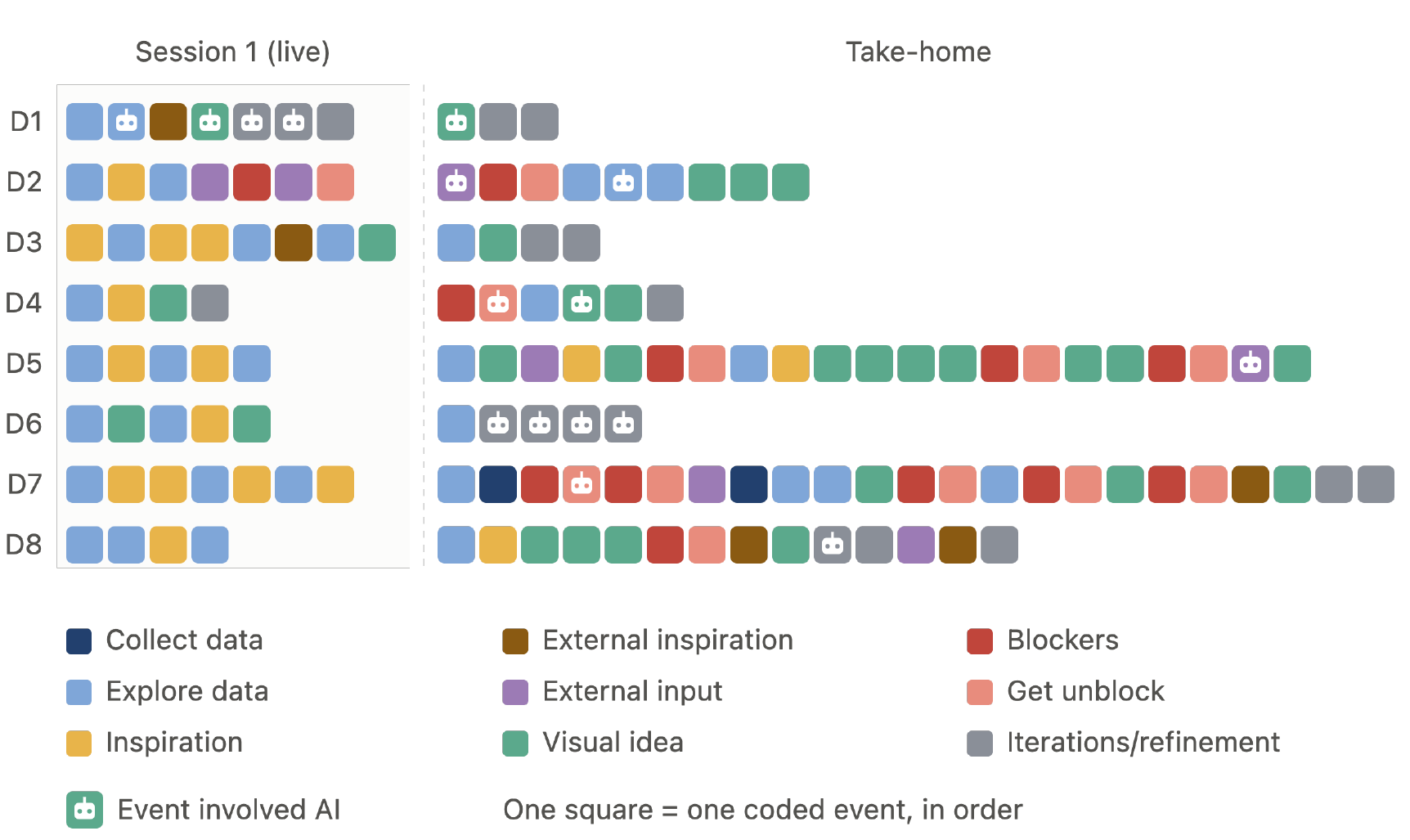}
  \caption{\textbf{Coded design-process sequences for the eight design-task participants (D1–D8).}}
  \Description{Rows of colored squares show each participant's activities during the live and take-home phases, with robot icons marking AI use.}
  \label{fig:S3_swimlane}
\end{figure*}

\subsection{Study III: Two-Session Design Task}

\begin{table*}[t]
\centering
\caption{\textbf{Study III participant demographics and practice.} We recruited 8 participants with diverse backgrounds in data visualization, design, and art.}
\Description{Eight rows describe participants D1 to D8 by sex, age range, and experience.}
\label{tab:S3_designers}
\small
\begin{tabular}{@{}lllp{0.8\textwidth}@{}}
\toprule
\textbf{ID} & \textbf{Sex} & \textbf{Age} & \textbf{Experience} \\
\midrule
D1 & Female & 25--34 & Data visualization designer; extensive experience in creative data visualization \\
D2 & Female & 25--34 & PhD student; academic data visualization; 20+ years of drawing and painting as a hobby \\
D3 & Female & 25--34 & Art educator; self-data visualization; qualitative self-tracking and weaving-based art \\
D4 & Female & 25--34 & PhD student; academic data visualization; painting experience \\
D5 & Female & 55+ & Business owner; 20+ years of professional data visualization; 10+ years of mixed-media and painting practice \\
D6 & Female & 25--34 & Tableau developer; Dashboard development; some experience in creative data visualization \& graphic design \\
D7 & Female & 18--24 & Research manager; experience in data and design; some experience in creative data visualization \\
D8 & Male & 25--34 & Graduate student; experience in data visualization; professional in UX and graphic design \\
\bottomrule
\end{tabular}
\end{table*}

\subsubsection{Participants}
We recruited through social media, practitioner communities including the Data Visualization Society and Tableau user groups, and by targeted email. To take part, a designer had to have experience creating data visualizations, and a background or experience in fine art, graphic design, interactive design, or creative data visualization. We limit participation to residents of the United States, for our institution's payment process constraint, and participants received \$70 on completion. Our university's Institutional Review Board (IRB) approved the study.

Twelve designers took part initially. Eight completed the full study; the remaining four did not return after the first session because they could not find time to work on it. We analyse data only from the eight who completed, and refer to them as D1--D8 (Table~\ref{tab:S3_designers}).

\subsubsection{Procedure}
The study ran across two video sessions with a period of independent work in between. Designers began with a dataset they had not worked on before. Both sessions were held over Zoom and screen-recorded.

\paragraph{Session 1 (40--45 minutes).}
We began by showing several examples of creative and artistic data visualization to establish what we meant by expressive work as distinct from a standard bar or line chart. Participants then chose one of six public datasets we prepared: solar and lunar eclipses, the MTA permanent art catalog, World Cup stadiums, AKC dog breed traits, Christmas songs on the Billboard charts, and endangered languages. The datasets ran from a few hundred to a few thousand rows. Participants were allowed to bring a dataset of their own provided they had not yet begun working with it, and none did. We told them that they have complete freedom deciding which part of a dataset to use since it's part of the idea. The final form was open and no finished piece was expected, only enough by the end of the study for us to picture what the work would be. Participants then spent around twenty minutes developing ideas in whatever way felt natural to them, sharing their screen and thinking aloud throughout.

\paragraph{Independent work.}
Participants continued working on it at their own pace after the first session. We asked participants to keep everything they produced, including rough, unfinished, and abandoned material. To support this we gave each participant a log card at the end of Session~1 and asked them to document whenever something notable happened such as a new idea, being stuck, an inspiration, a direction rejected, etc. 

\paragraph{Session 2 (30--40 minutes).}
Participants walked us through what they had produced since the first session. We then asked some follow-up questions about where ideas had come from, what they had rejected and why, where they had become stuck, whether they had considered alternative ideas, and what an ideal tool should be like to support them in the ideation phase. 

\subsubsection{Analysis}
We analyzed each participant's case in two stages. First, two authors conducted a close reading of the complete record---the Session~1 screen recording and think-aloud transcript, artefacts and log cards from the at-home design period, and the Session~2 transcript---to reconstruct how the project unfolded and to identify case-level observations and patterns across sources. This close reading was analytic in its own right and was not limited to preparing for coding. 
Second, we coded the reconstructions using the Study~I codebook unchanged, yielding eight event chains comprising 132 events (Figure~\ref{fig:S3_swimlane}). For each event, we additionally marked its phase and whether generative AI was involved. We treated AI involvement as an attribute rather than a separate code because participants used AI in roles---an expert to consult, someone to talk to, a way of looking something up, an assistant that writes code---already captured by the codebook, and a single AI code would have obscured these distinctions. Our findings draw on both the case-level observations and the coded process patterns. Finally, we compared these findings with Study~I and Study~II to identify convergence, divergence, and findings unique to this study.

\begin{figure*}[t]
  \centering
  \includegraphics[width=\textwidth]{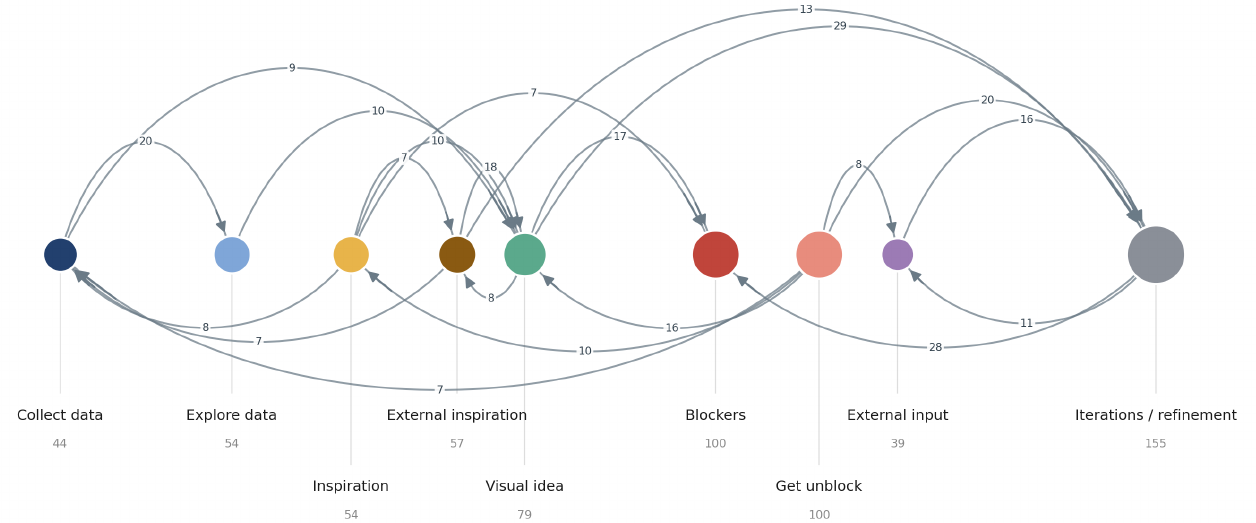}
  \caption{\textbf{How projects move from one activity to another.} Nine of the 11 top-level codes describe sequential project activities and are shown here. Each circle represents one activity, positioned by its median normalised location in the project process. Circle size reflects how often the activity occurs, with the total event count across the 40 articles shown below each label. Arrows show transitions between activities that occurred one after another: the arrow points to the next activity, and its label gives the number of observed transitions. To highlight common patterns, we show only transitions that appeared in at least 7 of the 40 articles. These 21 transitions account for 66\% of all transitions between different activities. Transitions between \textit{Blockers} and \textit{Get unblock} are omitted because every blocker was coded together with the response that resolved it.}
  \Description{Nine colored circles represent design activities, with curved arrows showing transitions and returns between them.}
  \label{fig:process_mining}
\end{figure*}
\section{Findings}
Our analysis across the three studies identified several recurring patterns in how creative and artistic data visualization design unfolds in practice. Rather than presenting the studies separately, we synthesize their findings around seven themes.

\subsection{The Overall Design Process}
\label{sec:overall}
\begin{figure*}[t]
  \centering
  \includegraphics[width=\textwidth]{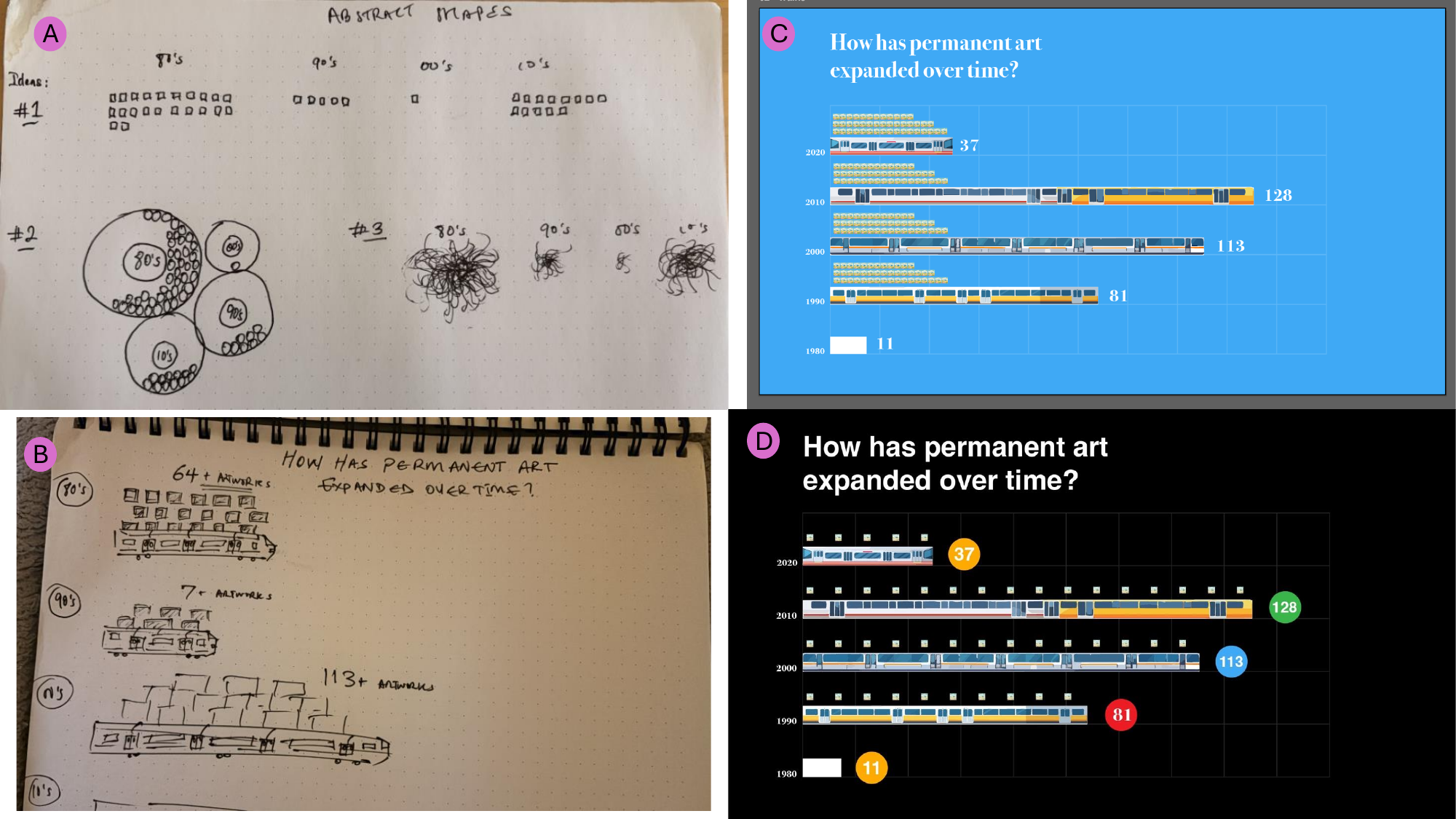}
  \caption{\textbf{D8's artifact changes during the design process.} (A) initial sketches exploring three visual ideas (Idea 1: simply squares; Idea 2: round container with other circles in there; Idea 3: thread-like thing to show the complexity of the train network); (B) a new sketch that was eventually adopted; (C) Illustrator refinement based on the new sketch; (D) further refinement after external inspiration}
  \Description{Four panels show D8's artifacts created during his design process.}
  \label{fig:design_process_D8}
\end{figure*}

\paragraph{The process follows a recognizable trajectory, but not a fixed route.}
Across the corpus and the observed design sessions, visualization design commonly moved from working with data and seeking inspiration toward a visual direction, then into making, encountering problems, resolving them, and continuing to refine the design. Study~I provides the broadest view of this trajectory: ordering the coded activities by their median position gives the average shape of a project, while Figure~\ref{fig:process_mining} shows the activities and transitions underlying it. In 87 of 260 transitions between different activities (33.5\%), designers returned to an activity that appeared earlier in the average process, and 35 of 40 write-ups contained at least one such return. These returns most often brought designers back to inspiration, data, or other people: from refinement to \textit{External input} (9 accounts), from a \textit{Visual idea} to \textit{External inspiration} (8), and from inspiration to \textit{Collect data} (8 and 7).
Study~III revealed a similar trajectory in the eight coded sequences (Figure~\ref{fig:S3_swimlane}). D8, for example, explored the MTA art catalogue to tell a story about the development of artwork over time. He rejected three initial sketches (Figure~\ref{fig:design_process_D8}A), then adopted a subway-train metaphor for artwork counts (Figure~\ref{fig:design_process_D8}B), and refined it in Illustrator (Figure~\ref{fig:design_process_D8}C). A conversation with a New York resident and searches for culturally relevant images prompted further refinement (Figure~\ref{fig:design_process_D8}D).

\paragraph{Designers describe this trajectory as flexible rather than mandatory.}
Study~II revealed the same structure from the designers' own accounts. I4 was the only participant who described her process as an explicit sequence: \textit{``idea, data, sketching, prototyping, making the colors beautiful, then legends and annotations.''} Yet she sometimes skipped sketching or prototyping when an idea was already clear, and returned to sketching when a prototype failed: \textit{``in your head it looks beautiful but with real data it looks awful.''}

\paragraph{Project format and goal do not strongly determine the overall process chain.}
Comparing projects by \textit{output format} and \textit{overall visualization goal}, we find little evidence that either systematically shapes the process chain (Table~\ref{tab:groups}). The projects intended to evoke attention and discussion are the only grouping that shows a significant difference before correction, but the effect is small and does not remain significant after correction. Given the small group sizes and the number of comparisons, these analyses are sensitive mainly to large differences.

\begin{table*}[t]
\centering
\caption{Comparison of process chains across projects grouped by output format and overall goal. We compare groups using three measures: the whole sequence, which captures both the activities used and their positions; the order of activities, which counts a run of the same activity once; and composition, which ignores order. $R^2$ is the share of between-project differences explained by the grouping, and $p$ is obtained from 9999 random reassignments of projects to groups. Across the 18 comparisons, only the \textit{evoke attention and discussion} grouping has $p<.05$; after correction, its value is $q=.12$.}
\Description{Rows compare project groups defined by output format or visualization goal. Paired columns report R-squared and p-values for whole sequences, activity order, and composition.}
\label{tab:groups}
\small
\setlength{\tabcolsep}{5pt}
\begin{tabular}{@{}llcccccc@{}}
\toprule
 & & \multicolumn{2}{c}{Whole sequence} & \multicolumn{2}{c}{Order of activities}
   & \multicolumn{2}{c}{Composition} \\
\cmidrule(lr){3-4}\cmidrule(lr){5-6}\cmidrule(lr){7-8}
Attribute & Groups compared & $R^2$ & $p$ & $R^2$ & $p$ & $R^2$ & $p$ \\
\midrule
Output format
  & static (18) vs.\ screen (11) vs.\ physical (11)          & .058 & .33 & .064 & .19 & .052 & .40 \\
  & physical or installation (11) vs.\ rest (29)       & .025 & .51 & .025 & .50 & .015 & .72 \\
\addlinespace
Visualization Goal
  & evoke attention and discussion (14) vs.\ rest (26) & .056 & \textbf{.013} & .056 & \textbf{.011} & .058 & .05 \\
  & personal interest (11) vs.\ rest (29)              & .029 & .36 & .025 & .49 & .022 & .48 \\
  & exhibition or open call (11) vs.\ rest (29)        & .030 & .33 & .032 & .25 & .011 & .83 \\
  & engaging to gain insights (7) vs.\ rest (33)       & .015 & .83 & .042 & .08 & .039 & .17 \\
\bottomrule
\end{tabular}
\end{table*}

\subsection{Sources of Inspiration}
\label{sec:inspiration}
\paragraph{Designers often begin by deciding what the piece should be about.}
Design ideas came from diverse sources, but designers often began by figuring out what the piece should be about rather than what it should look like. In Study~III, every participant first considered a story or subject. D4 described this as her usual starting point: \textit{when I work with a dataset, what I'll do first \ldots I want to figure out, like, what story I want to tell.''} D5 immediately noted the absence of such a goal in the context of our study: \textit{typically, I want to have an objective that I'm working towards when I start. And in this case, it's kind of weird, because we don't have an objective.''} All eight participants recorded initial ideas in their own ways, using paper, digital notes, or the tools they were already working with. For example, D5 and D8 recorded ideas directly in Tableau while exploring the data.

\paragraph{Finding a story and exploring the data are intertwined.}
Looking for that story was not a separate phase from data exploration. Figure~\ref{fig:S3_swimlane} shows that transitions between \textit{Explore data} and \textit{Inspiration} account for over half of all transitions between different codes, and five participants made this round trip more than once. D5 moved from sorting endangered languages by speaker count to exploring countries and felt she had \textit{``a beginning of a story.''} She then built and abandoned a map, narrowed the scope to the United States, and abandoned that framing too before the live session ended. By contrast, the corpus shows almost none of this back-and-forth. Across its 642 adjacent pairs, \textit{Explore data} is followed by \textit{Inspiration} only five times, and \textit{Inspiration} is followed by \textit{Explore data} four times, with the latter occurring in only three of the 40 accounts---neither transition is frequent enough to appear in Figure~\ref{fig:process_mining}. We interpret this difference as a property of what written accounts preserve rather than how the two groups work. A finished narrative tends to preserve the idea that survived and the data work that supported it, while leaving out the small returns to the data, abandoned directions, and intermediate explorations that helped produce it.

\begin{figure*}[t]
  \centering
  \includegraphics[width=\textwidth]{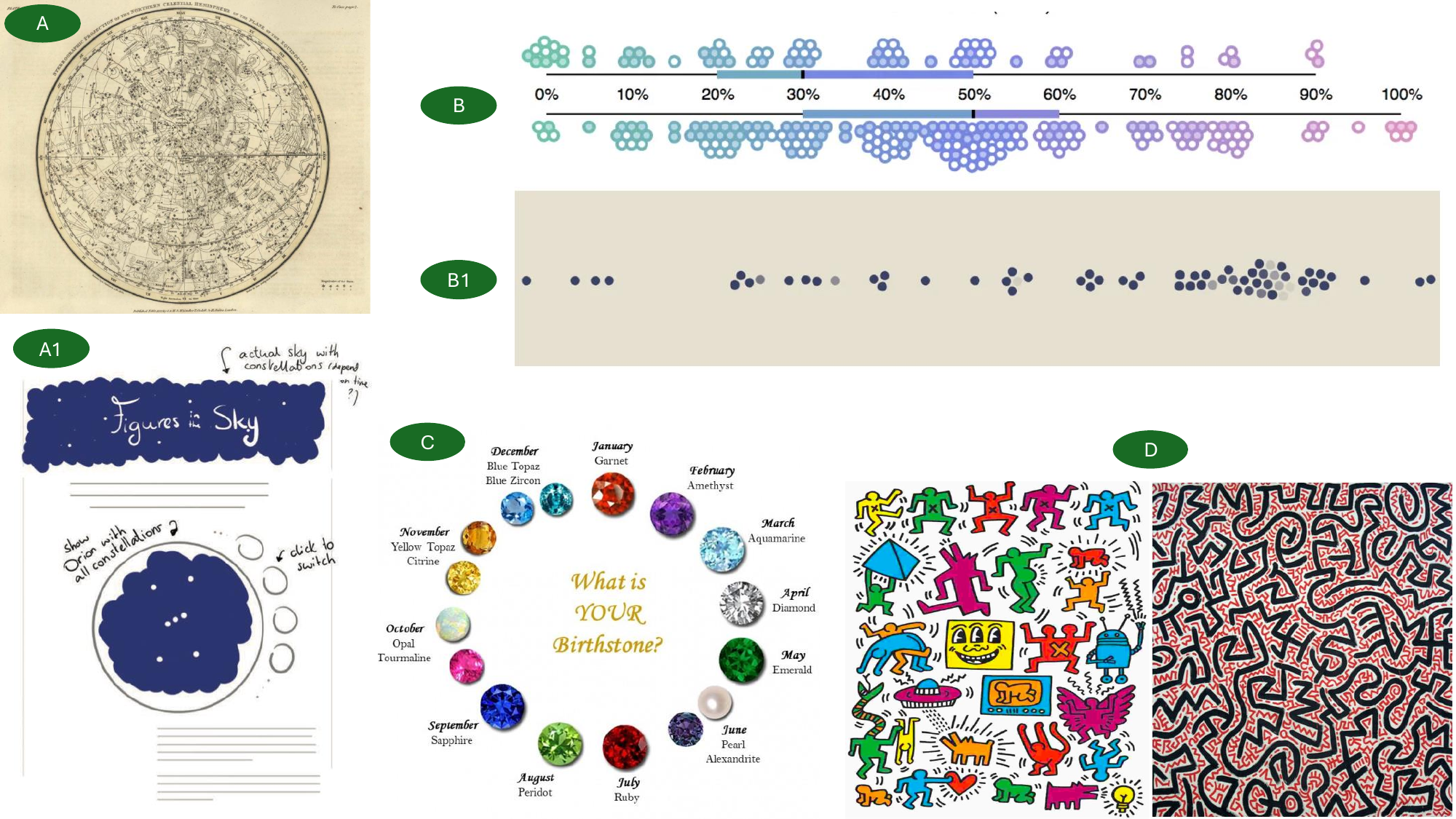}
  \caption{\textbf{Examples of inspiration sources from Study~I (A/A1 and B/B1) and Study~III (C and D)}}
  \Description{A collage of inspiration sources and adapted designs, including sky maps, dot charts, birthstones, and colorful outlined artwork.}
  \label{fig:inspirations}
\end{figure*}

\paragraph{Visual ideas come from both external inspiration and the data itself.}
In Study~I, \textit{External inspiration} was the most common activity immediately preceding a \textit{Visual idea}, accounting for 23.1\% of such transitions; after seeking external inspiration, authors moved to a \textit{Visual idea} 32\% of the time. Most external inspiration involved other people's work: \textit{Others' work} accounted for 71.9\% of external-inspiration events and appeared in 27 of 40 accounts. For example, in Nadieh's \textit{Figures in the Sky}~\cite{bremer2021datasketches}, she first looked at ancient sky maps (Figure~\ref{fig:inspirations}A) for the star map and then sketched the general layout (Figure~\ref{fig:inspirations}A1). In \textit{Dear America}~\cite{stahl2020foreigninterference}, the author was searching for an overarching way to present foreign-interference cases when he recalled Shirley Wu's \textit{655 Frustrations Doing Data Visualization}~\cite{wu2017frustrations} and redesigned his sketch (Figure~\ref{fig:inspirations}B1) around Wu's model (Figure~\ref{fig:inspirations}B). Study~II interviewees described similar routines. I6 starts a Pinterest board for a new dataset to collect \textit{``colors, themes, design elements.''} Sometimes everyday objects also inspired forms: : I6 drew inspiration from a sweater for a piece about a Christmas film, while I7 found an encoding for a bird count in an IKEA curtain.

Data exploration was another route to visual form. In the corpus, \textit{Collect data} and \textit{Explore data} together accounted for 24.4\% of the activities preceding a visual idea. I1 finds forms by pushing data through his tools and watching what emerges: \textit{I don't know what it looks like---I've got to see the shape. So this is just trying random ideas to see what shapes start looking interesting.''} Kirell Benzi (I5) similarly traced the inspiration for a year-long project, HPC Data Symphonies\footnote{\url{https://kirellbenzi.com/work/hpc-data-symphonies/}}, to \textit{successively trying to iterate,''} with processing choices ruling out options before anything was drawn: \textit{``You process it and see that one dimension is interesting but 70\% is missing, so you already know it won't work.''}

\paragraph{Designers often adapt existing visual forms rather than inventing everything from scratch.}
I4 adapted folk-painting symbols for a family tree because \textit{they already have symbols with meaning.''} while I7 used Van Gogh's \textit{Starry Night} as the ground of a piece and checked whether its elements could accommodate the data. In Study~III, D3 encoded months using birthstones (Figure~\ref{fig:inspirations}C), and returned to a small-weaving format she had used before. D7 borrowed Keith Haring's style as \textit{``an easy way to signal `New York subway' without explicitly including a subway in the final artwork''} (Figure~\ref{fig:inspirations}D).

\paragraph{Inspiration is difficult to explain.}
When an idea arrived, designers could rarely say exactly why. I7 explained, \textit{``I don't know why my brain thought that \ldots sometimes you make connections without even noticing.''} Her strategy was to create conditions in which a connection might emerge---go to a gallery, go for a walk, and wait---and she noted that the waiting can last months.

\subsection{Stages Define Problems and Responses}
\label{sec:stages}
\paragraph{The problems designers encounter change as the project progresses.}
In the corpus, early problems are more often about what is missing or not yet settled, such as \textit{Missing data}, \textit{Uninformative data or analysis}, and \textit{No visual direction}, with median positions of 0.12, 0.15, and 0.24, respectively.
Later problems shift toward the emerging artifact itself, including an \textit{Unsatisfying visual}, \textit{Technology failed}, \textit{Material or fabrication limitation}, or \textit{Too much data} (at 0.57, 0.58, 0.61, 0.73, respectively). 

\paragraph{The stage of a problem shapes how designers respond to it.}
Earlier responses are more likely to change direction, such as \textit{Abandon the direction} (median position 0.23), \textit{Reframe the problem} (0.25), and \textit{Subjective data filter} (0.26). Later responses more often adjust an existing direction, including \textit{Change the material or technique} (0.64), \textit{Simplify or reduce scope} (0.73), and \textit{Put aside} (0.76). Among the 11 blocker types that received both kinds of responses, responses that changed direction occurred earlier for 9. Thus, the same difficulty is more likely to prompt a change in direction when it appears early and an adjustment to the existing design when it appears later. In the \textit{Compliment}~\cite{marie2024visualcompliment} project, for instance, the designer abandoned an entire butterfly concept when the first sketches looked lopsided (position 0.20), but addressed the same dissatisfaction later by swapping the symbols inside an established composition for flowers (0.70).

\paragraph{Outside help also shifts from building knowledge to evaluating the work.}
\textit{External input} appears in 21 of 40 write-ups. Authors tend to seek knowledge or skills earlier in the process (\textit{Seek out knowledge or skills}, median position 0.41; 18 events in 15 accounts), while input from friends, colleagues, or the public appears later (\textit{Friends/families/colleagues/public}, 0.71; 11 events in 9 accounts). 

\subsection{Roles of Real Data in Visual Representation}
\label{sec:realdata}
\paragraph{Sketching helps designers externalize an idea, but does not establish whether it works.}
Six of seven interviewees had sketched in their projects, but sketching did not establish how an idea would work with real data. I1, who did not sketch, called the dataset \textit{the source of truth.''} In one piece, he mapped note length to circle size, but tied notes exported as a single long note produced a circle that dominated the image. Benzi emphasized that real data constrains aesthetic choices: \textit{``Real data is messy \ldots I think this is what sets data art apart from generative design or other art, where you can maximize pure aesthetics.''}

\paragraph{Sketches and prototypes support different kinds of work.} Corpus analysis reveals that the two forms of externalization appear with similar frequency: sketches appear 32 times across 23 articles, while prototypes appear 33 times across 32 articles. However, the difference between them is in what happens next. After a sketch, the next thing is refinement 45\% of the time and a blocker 16\% of the time; after a mockup, refinment is 31\% of the time and blocker is 25\%. The refinement differs too: 8 of the 14 refinements following a sketch are about \textit{Mapping}, while 6 of the 10 following a mockup are about the \textit{Visual}. Sketches are therefore where the encoding gets worked out, while mockups made with real data are where problems become visible.

\paragraph{Testing ideas on real data can rule out or redirect them before substantial production.} 
In the design task, this test happened early and often. D2 arrived with a concept---artworks fading in and out along a timeline of stations---and spent her first minutes sorting by station and line to see whether the artworks changed at all, concluding that it \textit{``might just be really, honestly, sparse data for me to visualize in the way I was thinking.''} Every participant ruled out fields and some ruled out whole concepts, for reasons including sparsity, lack of interest, insufficient information, illegibility, excessive literalness, or skill requirements. Missing data could redirect the design as well: four participants wanted a field the dataset did not contain, and each adapted the design in response. In Study~I, missing data appears as a blocker early in an account.

\paragraph{Working with real data also carries a production cost.}
Hands-on data work costs interest and enthusiasm as well as time. I3 avoided an idea that required building one tile and then replicating it in Illustrator \textit{``because I was afraid I would put too much effort into it, and that I'd get tired of this project.''} I7 entered more than two hundred coordinates into Tableau one point at a time, which she called \textit{``very frustrating''} and, in French, \textit{``\emph{travail de fourmi}, ant's work.''}

\subsection{Blockers and Getting Unblocked}
\label{sec:blocker-response}

\paragraph{The medium constrains which ideas are feasible in the first place.}
D3 filtered ideas continuously through what she knew about weaving: the warp is set once for each piece, so anything that varies between pieces has to be carried by the weft, and \textit{``24 is a nice weaving number''} suggested that a decade of eclipses would divide well. The one thing she could not settle during our study was the weave structure, which \textit{``would take experimentation and looking at the weaving literature.''} The medium also affected which problems occurred at all. In the corpus, a \textit{Material or fabrication limitation} appeared in 6 of the 11 physical projects but only 2 of the other 29 ($p = .003$, $q = .13$), while an \textit{Unsatisfying visual} appeared in none of the 11 physical projects and in 9 of the other 29 ($p = .043$, $q = .65$).

\begin{figure*}[t]
  \centering
  \includegraphics[width=\textwidth]{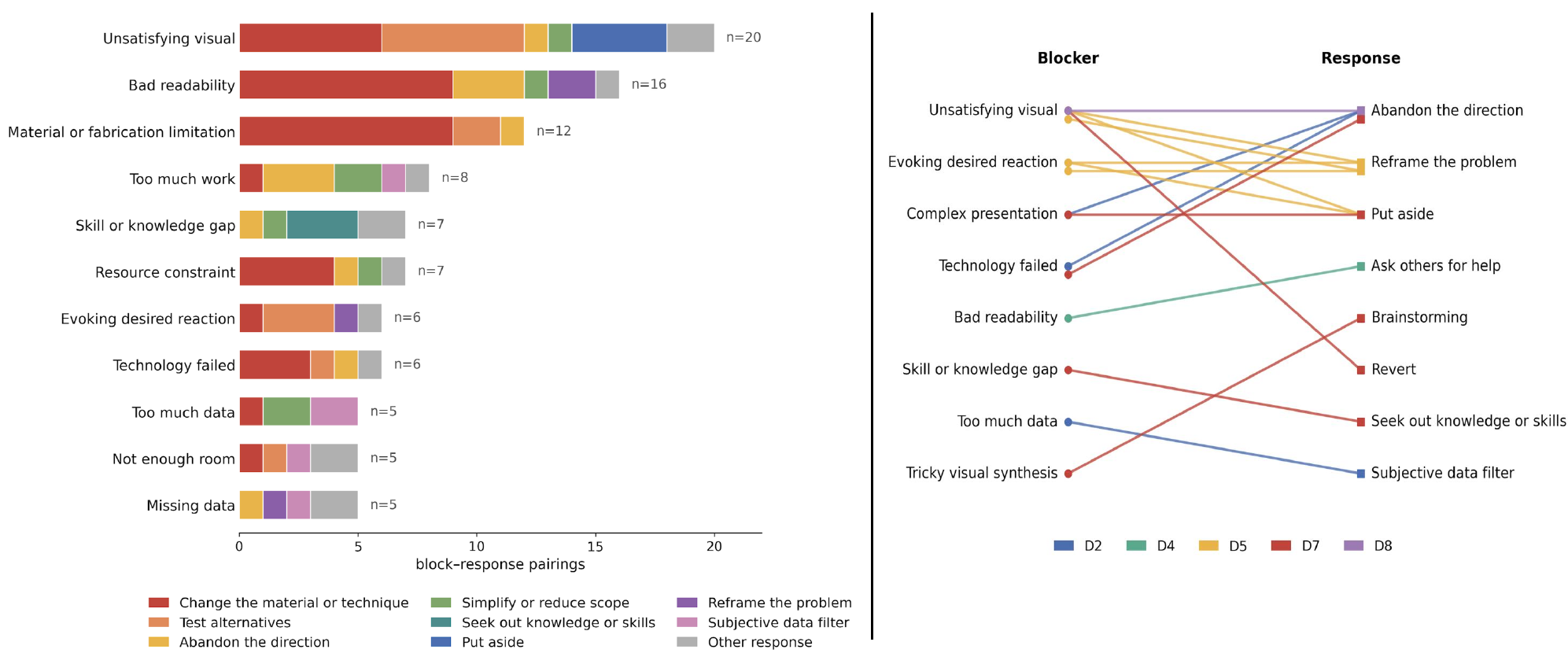}
  \caption{\textbf{Blocker-response pairs.} Left: Study~I. Bar length indicates the number of pairings. Blocker types with fewer than five pairings are omitted. Response types are shown separately only when they occur in at least five pairings; rarer responses are grouped as \textit{Other response}. Right: Study~III. Each blocker-response pair is shown in the participant's color.}
  \Description{Stacked bars on the left and connecting lines on the right show relationships between design blockers and responses in two studies.}
  \label{fig:blocker_pairs}
\end{figure*}

\paragraph{Designers rely on a relatively small set of responses to blockers, but the best response depends on the problem.}
Figure~\ref{fig:blocker_pairs} (Left) shows the matching of blockers with their responses. \textit{Change the material or technique} accounts for 34\% of the 100 \textit{Get unblock} events. The response often depends on the type of problem: 9 of the 12 responses to a \textit{Material or fabrication limitation}, and 9 of the 16 responses to \textit{Bad readability}, involve changing the material or technique. By contrast, dissatisfaction with the visual form leads to a wider range of responses: the 20 responses span seven different sub-codes, with none accounting for more than a third. In the design task (Figure~\ref{fig:blocker_pairs} (Right)), responses are mainly concentrated in \textit{Abandon the direction}, \textit{Reframe the problem}, and \textit{Put aside}.

\paragraph{Readability places a boundary on how far designers can push creative interpretation.}
Benzi built a background from an L-system and liked how it looked, but because the shape came from a grammar rather than the data, \textit{``no one would recognize the data in that crazy 3D shape,''} so he returned to something simpler. I7 gave up brush-stroke marks because three sizes were too difficult to distinguish. I6 stated the constraint directly: \textit{``you never want to go too far into a creative interpretation, because then the data might get lost.''} 
 
\paragraph{Pausing can change how designers see an existing idea rather than generate a new one.}
I6 shelved a piece because a handwritten style \textit{``felt really messy to me,''} but three months later the same quality read differently: \textit{``the messiness is part of the fun. And so I got over it.''} Sometimes an old attempt became useful only after returning to it, as when I1 came back, \textit{``looked at this earlier version, and thought, maybe I can do something with that.''}
 
\paragraph{Designers stop when the artifact feels sufficiently successful rather than when it is objectively complete.}
Design-task participants stopped when the piece met their communicative or aesthetic aims. D1 asked whether anything could be removed without losing the point; D3 stopped before adding information made the piece visually \textit{ugly''}; and D5 sought alignment between headline, copy, and form. Interviewees also relied on intuition: I3 described it as \textit{``something just clicked, it felt right \ldots gut feeling, intuition, I don't know exactly how to name it.''}

\subsection{Sequential but Iterative}
\label{sec:iteration}
\paragraph{Designers spend most of the process refining an established direction.}
\textit{Iterations/refinement} accounts for 155 of 682 events (22.7\%) and appears in 38 of 40 articles in Study~I. The commonest three-step pattern is \textit{Collect data} $\rightarrow$ \textit{Visual idea} $\rightarrow$ \textit{Iterations/refinement}, in 31 of the 40 accounts. Once refinement begins, authors rarely return to data work: among the 38 articles that reach refinement, only 5 return to \textit{Collect data} or \textit{Explore data}, compared with 18 that return to inspiration or another \textit{Visual idea}. 

\paragraph{Iteration usually happens around one active direction rather than through parallel development of complete alternatives.}
Interviewees typically developed one candidate at a time. I7 explained, \textit{``usually my brain doesn't go in 10 different directions \ldots if there is one particular one, I will make it work.''} I4 keeps alternatives on paper but usually pursued one first: \textit{``if it's not okay, return to the idea, but if it's okay, follow the first idea.''}

\paragraph{The cost of producing a candidate helps explain this convergence.}
One direction could mean six months of assembling a material piece (I4), or a folder of a hundred files (I1). I3 abandoned a promising sketch \textit{``because I thought I wouldn't have time to pursue it,''} and Benzi's HPC Data Symphonies project ran for a year, with four or five months of it active. We did not ask for a finished visualization in Study~III; while some participants carried out iterations and refinements, they all proceeded in a single visual format.

\subsection{Use of AI}
\label{sec:ai}
\begin{figure*}[t]
  \centering
  \includegraphics[width=\textwidth]{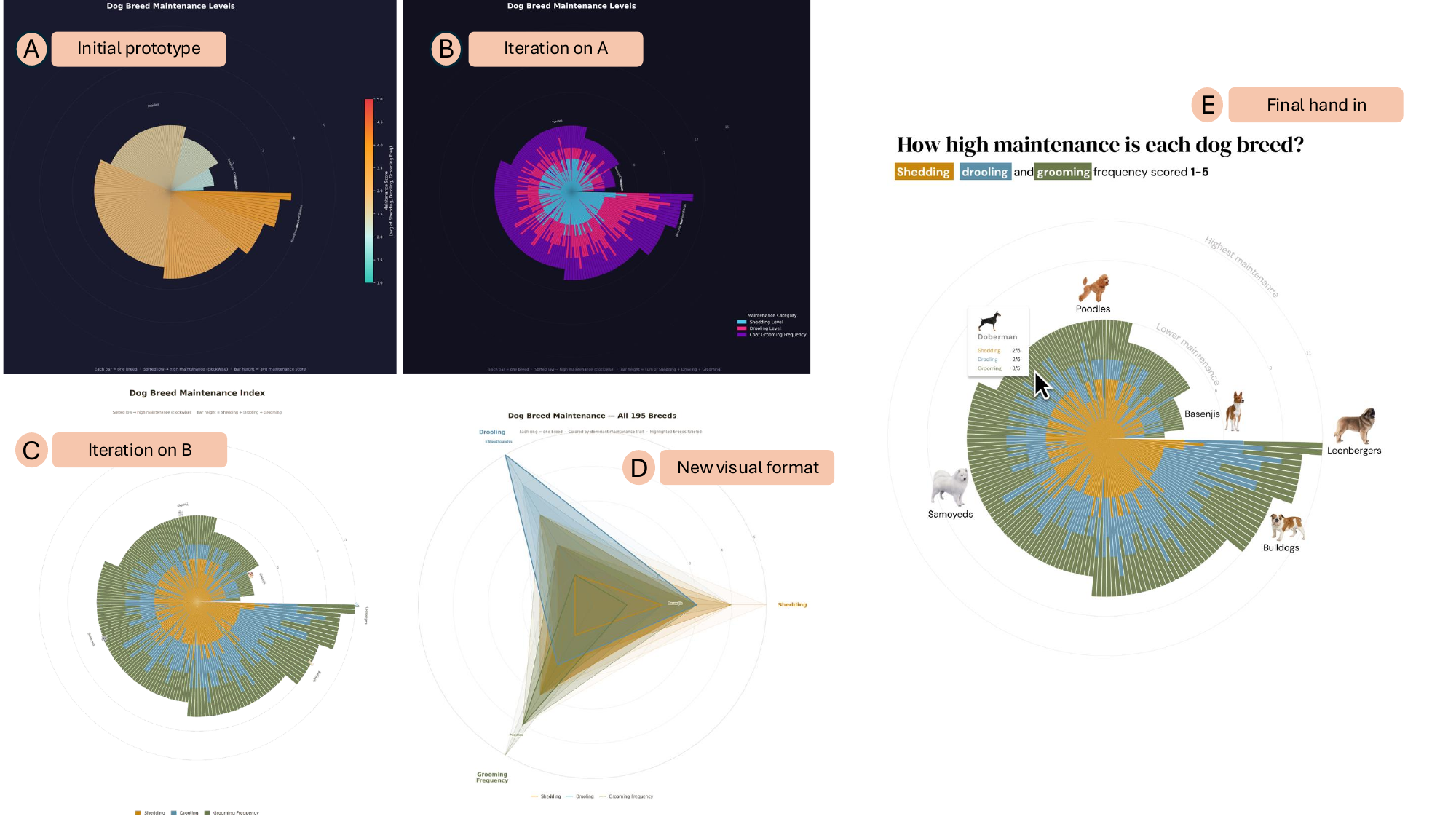}
  \caption{\textbf{D1's prototype changes with AI assistance.} Panels A-D were produced by Claude and E by D1 by hand. Claude built an initial radial chart from D1's data, example chart, and headline (A), regenerated it to fix how categories were shown (B), and restyled it (C). A radar chart made for comparison (D) was set aside. D1 then finished the radial design from C manually in Figma (E).}
  \Description{Five panels show D1's AI-assisted design iterations.}
  \label{fig:AI_use_D1}
\end{figure*}

\paragraph{AI was absent from the historical corpus but common in the live design task.}
Not one of the 40 articles mentions AI anywhere in a design process. We take this as a fact about the years those accounts cover rather than about practice today. In the design task, seven participants used a generative model. D1 used Claude to turn her dataset, reference chart, and headline into a prototype (Figure~\ref{fig:AI_use_D1}A), correct a category-display error (Figure~\ref{fig:AI_use_D1}B), and revise styling (Figure~\ref{fig:AI_use_D1}C). She also requested an alternative chart for comparison (Figure~\ref{fig:AI_use_D1}D), but retained the first prototype and refined it manually in Figma (Figure~\ref{fig:AI_use_D1}E).

\paragraph{AI served mainly as a prototype builder, data-exploration aid, outside perspective, or technical assistant.}
Across the design task, models were used for building prototypes and iterating on them (D1, D4, D6); exploring data and returning candidate outputs (D1, D4, D6); consulting about a direction the designer already had but was unsure about (D2, D4, D5); and ordinary technical assistance (D2, D7, D8).

\paragraph{AI reduced the cost of checking ideas without taking over the design judgment.}
Designers already sought outside perspectives: after repeatedly revising a legend, I4 learned from friends that it remained unclear, concluding that \textit{``you need a fresh view on it.''} I2 similarly noted that, as the creator, she notices details more readily than a new viewer. In the design task, AI could provide this additional perspective: D2, D4, and D5 brought uncertain directions to a model to see whether they held up, and D5 configured hers to \textit{``find blind spots and biases and play devil's advocate.''} AI also made alternative prototypes cheap enough to test. I6 wanted a tool that \textit{``gets what's in my head out before I actually start designing it,''} while D1, D4, and D6 had models build prototypes for them. D1 even produced a second chart type only for comparison and observed that \textit{``if I were doing this myself, I wouldn't have prototyped another chart.''} The model did not choose the idea or the final direction; it made another option cheap enough to build and evaluate.

\paragraph{The observed use of AI was also shaped by the task constraints.}
What we saw is not necessarily how these designers work in their usual practice. D6 would have built her piece in Tableau, since \textit{``the only reason I used Claude was limited time and bandwidth.''}

\section{Discussion}

\subsection{Situated Judgment and Exploration of Alternatives}
\label{sec:discussion-process}
Our findings reinforce Parsons's account of visualization practice as grounded in experience and situated judgment~\cite{design_practice}. Designers adapted their activities as projects developed, revisited earlier decisions, and judged when to stop through the relationship between visual appeal and intended communication (Section~\ref{sec:overall} and Section~\ref{sec:blocker-response}). These observations suggest that support for creative visualization should accommodate this flexibility, allowing designers to return to data, references, and earlier representations as their questions change. Such support would preserve access to unfinished directions without requiring every project to follow a prescribed sequence.

Our study also reveal practical constraints on exploring of alternatives. The Five Design-Sheet methodology encourages distinct sketches before realization~\cite{DBLP:journals/tvcg/RobertsHR16}, while Dow et al. demonstrated benefits of parallel prototyping in an advertisement design task~\cite{dow2010parallel}. In our studies, early consideration of several possibilities often gave way to sustained development of one direction, with alternatives revisited when progress stalled (Sections~\ref{sec:overall} and~\ref{sec:iteration}). This pattern does not establish whether parallel development would have improved the work. It highlights the effort required to make alternatives concrete enough to evaluate. 

Reported production costs and the observed use of AI to produce additional prototypes motivate lowering this effort (Sections~\ref{sec:realdata} and~\ref{sec:ai}). Support should help designers develop candidate representations with actual data, inspect their formats, and revise them before detailed production. Its value depends on whether examining alternatives informs decisions about what to retain, change, or pursue, rather than simply increasing the number of options generated.

\subsection{Inspiration Develops through References and Data Exploration}
\label{sec:discussion-inspiration}
Our findings support Baigelenov et al.'s account of inspiration drawn from existing visualizations, everyday phenomena, and personal experience, with sources adapted into new designs~\cite{viz_inspiration}. Our observations explain how this adaptation relates to a visualization's subject and audience (Section~\ref{sec:inspiration}). Familiar imagery and cultural symbols contributed recognizable meanings, while existing forms suggested ways to organize or encode data. A reference's usefulness therefore depended on its relevance to the project as well as its appearance.

Data exploration offers another route to inspiration, informing both to identify a subject and to investigate the expressive possibilities of the data (Section~\ref{sec:inspiration}). Designers examined what patterns might support a story, and how the data could take shape through different mappings and arrangements. The resulting forms could themselves make a direction interesting or prompt its revision. This suggests that tools for creative exploration should connect the search for meaningful data patterns with opportunities to experiment with visual form, allowing designers to develop what they want to communicate through trying different ways of representing it.

Participants externalized possible stories through informal notes, sometimes within the tools used for data exploration (Section~\ref{sec:inspiration}). This practice resonates with Stokes et al.'s account of written questions and intended takeaways as guides for visualization design~\cite{writing_rudders}.

\subsection{Prototyping with Real Data Helps Develop the Idea}
\label{sec:discussion-realization}
Sketching has an established role in visualization design~\cite{DBLP:journals/tvcg/RobertsHR16,walny2015sketching,keefe2008scientific}. Our findings reinforce its value for developing mappings and exploring possible visual forms (Section~\ref{sec:realdata}). However, the questions a sketch can address depend on how much data it represents. It may help assess an arrangement or metaphor, while leaving unclear how well it works with actual data values.

Our findings align with Bigelow et al.'s observations that real data can overturn designers' assumptions~\cite{bigelow2014reflections}, and with accounts of design attempts prompting changes in problem framing~\cite{parsons2025beyond}. Encountering data properties leads designers to revise encodings, narrow subjects, or reconsider directions (Sections~\ref{sec:realdata} and~\ref{sec:blocker-response}). Sparsity, missing values, and unexpectedly dominant marks affected what representations could express and how they appeared. Prototyping thus contributes to decisions about what to make as well as how to make it.

For designers, the practical implication is to choose what to externalize according to the question being investigated. Sketches can support comparisons of metaphors, arrangements, and mappings, while rough prototypes can make the relevant relationships visible before detailed styling. For physical work, material trials could similarly help investigate encodings alongside fabrication constraints identified in Section~\ref{sec:blocker-response}. Tools could facilitate movement between these forms of exploration. Connecting drawn elements to data would allow early ideas to be inspected without extensive manual production. Preserving the ability to redraw, alter mappings, and change data selections would keep prototypes useful for ideation as new questions arise, including when an experiment prompts a return to sketching.

\subsection{Design Implications for Supporting Creative Visualization}
\label{sec:discussion-tools}
The findings suggest several directions for supporting creative visualization design, from recording an emerging story to testing and revising its representation.

\paragraph{Support story development alongside data exploration.}
The repeated search for a story during data exploration (Section~\ref{sec:inspiration}) motivates ways to assess candidate ideas before investing substantial effort. Written notes or optional speech capture could retain possibilities as they arise, linked to the data selections and exploratory views that prompted them. Checking required fields, distributions, and subsets relevant to a proposed story would help designers judge whether the data can support it. Keeping these connections visible would make it easier to revise the subject or return to earlier ideas without reconstructing the analysis. The goal is to reduce effort spent pursuing infeasible directions and help designers converge on a story they find meaningful.

\paragraph{Connect references, sketches, and real data.}
Designers' use of external inspiration (Section~\ref{sec:inspiration}) suggests helping them gather material that could spark a visual idea. A temporary mood board could bring together chart examples, artworks, everyday photographs, and creative visualizations related to the dataset or story. From this collection, designers can compare references and adapt forms or metaphors on a sketching canvas. Access to selected columns, rows, and values would allow them to try these drawings with data. As in DataInk~\cite{xia2018dataink}, selected mark properties could be bound to values while others remain freely editable. Maintaining these connections and allowing further editing in preferred illustration tools would help designers move from an inspiring reference to a visual form that represents their data.

\paragraph{Reduce the cost of testing, comparing, and revising alternatives.}
Time-consuming, repetitive work can prevent a promising idea from receiving further attention (Sections~\ref{sec:realdata} and~\ref{sec:iteration}). Automating data entry, coordinate placement, and mark duplication could reduce this burden while preserving the manual operations designers use to explore a form. Easier production of rough prototypes would allow more alternatives to be examined before detailed work begins. Retaining versions, data selections, and mappings would support comparisons and returns to earlier ideas. Access to feedback on specific questions, such as whether a legend is understandable or a representation conveys its story, would give designers further grounds for revision.

\paragraph{Offer AI as optional assistance across these activities.}
AI could help assess written or spoken ideas by checking their data requirements and identifying relevant domain information. Explanations of why a direction lacks support, together with suggestions for narrowing or revising it, could help designers decide what to pursue. Alternative story angles and headlines could be offered on request. For inspiration, AI could retrieve suitable chart types, images related to the story, and creative visualizations suggesting useful forms or metaphors, with sources and explanations of their relevance. It could also generate editable prototypes and offer a possible reader's interpretation of a draft, identifying unclear stuffs. Designers should control when assistance is active and be able to inspect and correct its data transformations, mappings, and suggestions. All underlying activities should remain available without AI.

An integrated workspace connecting story notes, data, references, and visual drafts is one possible direction for future work. Evaluation of such a workspace could examine whether such connections reduce repetitive work, enable meaningful comparison of alternatives, and help designers understand and revise their decisions.

\subsection{Studying Inspiration as It Unfolds}
\label{sec:discussion-methods}
Our findings also inform research on visualization creation. Designers sometimes identified an inspiration source without explaining why it prompted an idea  (Section~\ref{sec:inspiration}). Asking for an explanation alone therefore gives limited access to how these connections arise. Process records allow researchers to examine what designers encountered, what they tried next, and how they later interpreted the connection. Even when the initial association remains unexplained, its use and development can still be studied.

The differences in how inspiration appeared across our studies (Section~\ref{sec:inspiration}) illustrate how different methods address different questions. Public blogs offer a broad view of acknowledged sources across projects, but selectively preserve what contributed to the finished work. Artifact-anchored interviews help explain why a source mattered and how it was adapted, drawing on memory and retained materials. Live observation captures brief explorations and discarded connections as they occur. Study~III combines observation with take-home cards and follow-up interviews, linking records of emerging ideas to designers' later explanations. Extending this approach to designers' own longer-term projects would help capture inspiration arising during pauses and encounters beyond scheduled design sessions.

\subsection{Limitations and Future Work}
\label{sec:discussion-limitations}

Studies~I and~II draw on publicly documented or self-selected completed projects, giving less coverage to brief explorations and abandoned work. Study~I's coding depends on narrative detail and researchers' judgments about activity boundaries and order. Accounts may compress repeated attempts or omit intermediate decisions, affecting which activities and transitions become visible. Event counts and positions therefore describe reported activities and their sequence, not the time or effort each stage actually took.

Study~III involved unfamiliar datasets without a client, fixed deadline, or requirement for polished or completed work. Four of the initial twelve participants withdrew because they lacked time, and five of the remaining eight chose the same dataset. These conditions limit coverage of how client requirements and delivery pressures shape ideation. Following ongoing professional projects could clarify their effects on story selection, iteration, and decisions to stop.

Finally, our understanding of how designers ideate for physical outputs comes mainly from the corpus, supplemented by two interviewees with relevant experience and one design-task participant pursuing a textile-based concept. We consequently have limited direct access to how material experimentation influences emerging ideas. Observing material trials and early samples could reveal how they generate possibilities or lead designers to revise a visual form.


\section{Conclusion}

Through 40 public project accounts, seven practitioner interviews, and a design task with eight participants, we examined ideation in creative and artistic data visualization. Ideas develop through data exploration, adaptation of references, and successive sketches and prototypes, with their trajectories shaped by production costs, practical constraints, and designers’ judgments. These findings motivate tools that connect stories, data, references, and visual drafts while making alternatives easier to test and revise. AI offers possible support within this process, but its benefits require further evaluation in ongoing design practice.




\bibliographystyle{ACM-Reference-Format}
\bibliography{references}

\appendix
\section{Codebook}
\label{app:codebook}
\begin{longtable}{@{}p{0.17\textwidth} p{0.23\textwidth} p{0.52\textwidth}@{}}
\caption{The full codebook: 11 top-level codes and 82 sub-codes.}
\label{tab:codebook} \\
\noalign{\Description{A three-column table groups sub-codes under their top-level categories and defines each sub-code.}}
\toprule
\textbf{Top-level code} & \textbf{Sub-code} & \textbf{Description} \\
\midrule
\endfirsthead
 
\multicolumn{3}{@{}l}{\textit{Table \thetable\ continued from the previous page.}} \\
\toprule
\textbf{Top-level code} & \textbf{Sub-code} & \textbf{Description} \\
\midrule
\endhead
 
\multicolumn{3}{r@{}}{\textit{Continued on the next page.}} \\
\endfoot
 
\bottomrule
\endlastfoot
 
Output format & Static 2D & A fixed image, in print or on screen. Includes an animated version produced from the same pipeline. \\*
\cmidrule(l){2-3}
 & Interactive 2D (screen) & The viewer controls the view on screen: zoom, pan, filter, hover, scroll. \\*
\cmidrule(l){2-3}
 & 3D on screen & Built from 3D geometry, whether delivered as fixed renders or as a navigable scene. \\*
\cmidrule(l){2-3}
 & Physical, non-interactive & A physical artifact the viewer looks at but does not operate. Sculpture, laser-cut object, woven or knitted piece, ceramics. \\*
\cmidrule(l){2-3}
 & Physical, interactive & A physical artifact or installation the viewer manipulates, or that responds to them. \\*
\cmidrule(l){2-3}
 & Multisensory /\allowbreak{} performance & Sound, taste, smell or movement as the primary channel. \\
\midrule
Overall visualization goal & Client's requirements & The purpose of the project is set by a client's or commissioner's brief. \\*
\cmidrule(l){2-3}
 & Evoke attention and discussion & The purpose is to draw attention to a subject and prompt reflection, debate, or awareness of it. \\*
\cmidrule(l){2-3}
 & Engaging to help gain insights & The purpose is to make the data engaging enough that a general audience takes an insight from it. \\*
\cmidrule(l){2-3}
 & Exhibition or open-call theme & The project is made for an exhibition, competition or open call, and that call's theme shapes the subject. \\*
\cmidrule(l){2-3}
 & Explore a new medium or technique & Part of the stated purpose is the author's own wish to work in a medium or technique they have not used before. \\*
\cmidrule(l){2-3}
 & Personal interest & The project exists for the author's own curiosity, practice or pleasure. The author is the main intended audience, along with the people around them. \\*
\cmidrule(l){2-3}
 & Coursework or research & The project exists to meet a course requirement, a thesis, or the aims of a research investigation. \\*
\cmidrule(l){2-3}
 & Inspire a particular reaction & The project is meant to make people feel a certain way, i.e. to invoke unease, joy, feel welcoming, etc. \\
\midrule
Blockers & Bad readability & The resulting visual cannot be read, or readability makes a particular visualization difficult. Includes occlusion, overdraw, overlapping marks and colours that cannot be told apart. \\*
\cmidrule(l){2-3}
 & Complex presentation & Complexity in representing something was a blocker or bottleneck \\*
\cmidrule(l){2-3}
 & Evoking desired reaction & The author has a specific reaction in mind that is difficult to evoke \\*
\cmidrule(l){2-3}
 & Missing data & Data the author needs is absent: missing from the dataset, or no data of adequate granularity exists for the intended subject. \\*
\cmidrule(l){2-3}
 & Not enough room & There is not enough space: in the visualization, or in the physical space the piece has to occupy. \\*
\cmidrule(l){2-3}
 & Technology failed & A technique, tool, component or piece of hardware did not work as needed, or the author judges in advance that it will not hold up. \\*
\cmidrule(l){2-3}
 & Too much data & Too much data to represent, fit, or analyse. Includes rendering and performance limits. \\*
\cmidrule(l){2-3}
 & Too much work & Something is too much work to do, or the amount of work is a bottleneck \\*
\cmidrule(l){2-3}
 & Tricky visual synthesis & Visually representing multiple items or datasets is challenging \\*
\cmidrule(l){2-3}
 & Unsatisfying visual & The author is not happy with the visual results, or the visuals fail to capture what the author aims to represent \\*
\cmidrule(l){2-3}
 & Skill or knowledge gap & The author says they lack a skill or knowledge that the idea requires. \\*
\cmidrule(l){2-3}
 & Material or fabrication limitation & The chosen material cannot do what the design needs, or making and assembling the artifact turns out to be difficult. \\*
\cmidrule(l){2-3}
 & Resource constraint & An externally fixed limit that the author has no power to change \\*
\cmidrule(l){2-3}
 & Uninformative data or analysis & The data and the analysis both work, but produce nothing worth showing. \\*
\cmidrule(l){2-3}
 & No visual direction & The author has no idea yet of what the piece should be, and says so. \\
\midrule
Collect data & Public & Existing records obtained from a public source. \\*
\cmidrule(l){2-3}
 & Clients & Data was given by a client. \\*
\cmidrule(l){2-3}
 & Self-collected/\allowbreak{}constructed & The author is the origin of the data: the records exist only because the author observed, asked, measured or recorded them. \\
\midrule
Explore data & Make basic charts & Author makes a standard bar/line/graph to help understand the data and see what's there \\*
\cmidrule(l){2-3}
 & Analysis & Data was analysed using data analysis methods, either qual or quant \\*
\cmidrule(l){2-3}
 & Annotation & Annotation was used as a part of exploring the data \\*
\cmidrule(l){2-3}
 & Data selection & A portion of the data was selected for representation \\*
\cmidrule(l){2-3}
 & Reading the data & The author reads the records or fields themselves to work out what the data contains, without computing anything. \\*
\cmidrule(l){2-3}
 & Transformed & A new dataset was created purely from the previous dataset \\*
\cmidrule(l){2-3}
 & Cleaning /\allowbreak{} preparation & Cleaning, restructuring, verifying or hand-assembling the data into a usable state. \\
\midrule
External input & Clients & Client provides data, design breif, or feedback. \\*
\cmidrule(l){2-3}
 & Expert & The author consults someone with relevant expertise: a domain expert, a teacher or mentor, a teaching assistant, a specialist in a community workshop, or the maintainer of a tool. \\*
\cmidrule(l){2-3}
 & Seek out knowledge or skills & The author goes and gets what they need to know or be able to do: searching online, consulting documentation, visiting a shop or library, and in its extended form taking a class or building a side project to acquire a capability. \\*
\cmidrule(l){2-3}
 & Friends/\allowbreak{}families/\allowbreak{}colleagues/\allowbreak{}public & The author gets input from people outside the project: friends, family, colleagues elsewhere, or a public call. \\
\midrule
External inspiration & Metaphor & The author draws visual inspiration from some metaphore \\*
\cmidrule(l){2-3}
 & Others' work & The author draws on something made by others: a specific work, a creator's body of work, a published technique, or a style or craft tradition. Encountered anywhere, online, in print, in a gallery, or in the literature. \\
\midrule
Inspiration & Brainstorming & The author thinks of many different ideas without necesarily trying all of them. \\*
\cmidrule(l){2-3}
 & Common data feature & Some common feature in the data (e.g. temporal information) acts as a inspiration for representation \\*
\cmidrule(l){2-3}
 & Eureka moment & A specific moment the author feels they have overcome a difficult blocker or have discovered the desired representation \\*
\cmidrule(l){2-3}
 & Happy little mistakes & A mistake turns into an inspiration \\*
\cmidrule(l){2-3}
 & Initial idea & The author has some initial idea, generally not specified where it comes from \\*
\cmidrule(l){2-3}
 & Obvious & The initial idea is 'obvious' \\*
\cmidrule(l){2-3}
 & Personal preference & The author has some personal preference from which they draw visual inspiration \\*
\cmidrule(l){2-3}
 & Physical properties & The author draws inspiration from physical properties of something \\*
\cmidrule(l){2-3}
 & Form seen in the artifact & The author recognises a shape or figure in an intermediate output and follows it. \\*
\cmidrule(l){2-3}
 & Affordance of the chosen medium & A medium or technology that has already been chosen suggests what to do with it. \\
\midrule
Get unblock & Put aside & The author gives up, walks away, or in some other way stops trying on a particular idea, possibly with the intention of returning to it later \\*
\cmidrule(l){2-3}
 & Abandon the direction & The author drops an idea, a dataset or a whole direction and moves to a different one. \\*
\cmidrule(l){2-3}
 & Pulling from former idea & The author brings in something from a former idea to solve a blocker \\*
\cmidrule(l){2-3}
 & Revert & The author returns to a previous version of the visualization or techniques to try again \\*
\cmidrule(l){2-3}
 & Subjective data filter & The author strategically cuts off part of the dataset to resolve a blocker \\*
\cmidrule(l){2-3}
 & Visualize data issues & The author decides to visualize a problem with the dataset \\*
\cmidrule(l){2-3}
 & Seek out knowledge or skills & The author goes and gets the knowledge or capability needed to clear a blocker, from a quick search up to a class or a side project. \\*
\cmidrule(l){2-3}
 & Transformed & A new dataset was created purely from the previous dataset in order to resolve a blocker. \\*
\cmidrule(l){2-3}
 & Brainstorming & The author generates several candidate solutions to a stated blocker. \\*
\cmidrule(l){2-3}
 & Ask others for help & The author approaches someone outside the project specifically to resolve a blocker. \\*
\cmidrule(l){2-3}
 & Change the material or technique & Swapping in a different material, component, methods, or technique, or adapting the existing one so that it works. \\*
\cmidrule(l){2-3}
 & Test alternatives & Author building or trys alternative options. \\*
\cmidrule(l){2-3}
 & Simplify or reduce scope & Making the plan smaller, simpler or partial so that it can proceed. \\*
\cmidrule(l){2-3}
 & Reframe the problem & The author changes the question rather than solving the one they started with, including letting go of a goal or constraint they set themselves. \\*
\cmidrule(l){2-3}
 & Get inspiration & The author gets external inspiration to try a new direction \\
\midrule
Iterations/\allowbreak{}refinement & Mapping & The author iterates on how data is connected to visual elements \\*
\cmidrule(l){2-3}
 & More data & More data was added to the visualization \\*
\cmidrule(l){2-3}
 & Not specified & Iteration is mentioned but what and how are not \\*
\cmidrule(l){2-3}
 & Visual & Any kind of visual iteration where different visual elements are tried \\*
\cmidrule(l){2-3}
 & Technical implementation & Work on how the piece is made or runs: code, rendering, performance, production files. \\*
\cmidrule(l){2-3}
 & Explanatory aids & Adding axes, legends, labels, a guide or an introductory element so that a viewer can read the piece. \\*
\cmidrule(l){2-3}
 & Interaction & Changing what the viewer can do, or how the piece responds to them. \\*
\cmidrule(l){2-3}
 & Material or medium choice & What the piece is made of or delivered in. \\
\midrule
Visual idea & Mockups & A produced draft of the visual. The tool is not part of the definition: design software and code both count, and accounts often do not say which was used. \\*
\cmidrule(l){2-3}
 & Sketching & Author sketches something with pen and paper or digital pen \\*
\cmidrule(l){2-3}
 & Described in words & The idea is put forward in words only; at this point in the account there is no sketch, draft or prototype. \\*
\cmidrule(l){2-3}
 & Physical prototype & The idea is externalised as a physical mock-up or test build. \\
\end{longtable}

\end{document}